\documentclass[notitlepage, nofootinbib,
final,
superscriptaddress,
groupedaddress,
unsortedaddress,
preprintnumbers,
 amsmath,amssymb,
 prd,onecolumn,12 point
]{revtex4-2}

\usepackage{graphicx}% Include figure files
\usepackage{dcolumn}% Align table columns on decimal point
\usepackage{bm}% bold math
\usepackage{hyperref}% add hypertext capabilities
\usepackage[mathlines]{lineno}% Enable numbering of text and display math
\usepackage{float}
\usepackage[T1]{fontenc}
\usepackage{color}
\usepackage{subfigure}
\DeclareUnicodeCharacter{2212}{\ensuremath{-}}

\begin{document}

%\preprint{APS/123-QED}

\title{A Common Origin for Diverse Neutrino Mass Matrix Textures}% Force line breaks with \\
%\thanks{A footnote to the article title}%

\author{Manash Dey}%
%\email{manashdey@gauhati.ac.in}
\email{manashdey1272@gmail.com}
\affiliation{Department of Physics, Maibang Degree College, India, 788831.}

\author{Pralay Chakraborty}%
\email{pralay@gauhati.ac.in}
\affiliation{Department of Physics, Gauhati University, India, 781014.}

\author{Dipshikha Das}
\email{d.das@iitg.ac.in}
\affiliation{Department of Physics, Indian Institute of Technology Guwahati, India, 781039.}

\author{Subhankar Roy}
\email{subhankar@gauhati.ac.in}
\affiliation{Department of Physics, Gauhati University, India, 781014.}%Lines break automatically or can be forced with \\

%\collaboration{MUSO Collaboration}%\noaffiliation

%\author{Charlie Author}
 %\homepage{http://www.Second.institution.edu/~Charlie.Author}
%\affiliation{
% Second institution and/or address\\
% This line break forced% with \\
%}%
%\affiliation{
% Third institution, the second for Charlie Author
%}%

\date{\today}% It is always \today, today,
             %  but any date may be explicitly specified

\begin{abstract}
We plan to decipher the common origin of neutrino mass textures within a $\Delta(27)$ based framework, in association with Type-I+II seesaw mechanism. Motivated by the diversity of texture structures used in phenomenological analyses, we examine how a single framework can reproduce them without additional assumptions. We show that the interplay between vacuum alignments, model parameters, and specific Yukawa choices naturally yields several well established neutrino mass textures in the literature. Our goal is to demonstrate that a minimal $\Delta(27)$ framework can unify these viable textures within a common theoretical origin.
\end{abstract}

%\keywords{Suggested keywords}%Use showkeys class option if keyword
                              %display desired
\maketitle
%\tableofcontents
\section{Introduction}
\label{sec:intro}
The discovery of neutrino oscillations~\cite{Pontecorvo:1957cp, Suzuki:1998rd, Super-Kamiokande:1998uiq, Super-Kamiokande:1998kpq, SNO:2002tuh, KamLAND:2002uet} firmly established that neutrinos\,\cite{Cowan:1956rrn, Davis:1968cp} have non-zero masses, providing the first clear indication of physics beyond the Standard Model (SM)\,\cite{Glashow:1961tr, Weinberg:1967tq, Salam:1968rm, Gross:1973id, Politzer:1973fx}, in which neutrinos are massless. This motivated several theoretical extensions of the SM that can account for the smallness of neutrino masses. Among these, the seesaw mechanisms are among the most widely studied and well motivated frameworks. Specifically, the Type-I seesaw\,\cite{Cai:2017mow, Mohapatra:2004zh, King:2013eh, Mohapatra:2006gs, King:2003jb} introduces heavy right handed neutrinos, while the Type-II seesaw~\cite{Melfo:2011nx, FileviezPerez:2008jbu, King:2003jb, Cheng:1980qt, Cai:2017mow} extends the scalar sector by incorporating an $SU(2)_L$ scalar triplet $\Delta$, which directly contributes to the Majorana mass of neutrinos.

In parallel, the observed structure of neutrino mixing led to significant interest in discrete flavour symmetries such as $A_4$\,\cite{Ma:2001dn, King:2006np, Altarelli:2010gt, King:2013eh}, $S_4$~\cite{Ma:2005pd,Hagedorn:2006ug}, $\Delta(27)$~\cite{Branco:1983tn, Ma:2007wu, CentellesChulia:2016fxr, Dey:2024ctx, Chakraborty:2025shb}, etc. These symmetries play the central role in shaping viable neutrino mass models: they not only restrict the allowed interactions, but also guide the structure of the neutrino mass matrix through specific vacuum alignments of scalar fields. This often leads to characteristic mixing patterns. A well known example is the $\mu$-$\tau$ symmetry\,\cite{Harrison:2002er}, which in its exact form predicts a maximal atmospheric mixing angle and a vanishing reactor angle. However, the experimental observation of a non-zero reactor angle required modifications to this simple picture, making discrete flavour symmetries even more crucial for constructing consistent models.  

The experimental data available today shows that several mysteries remain embedded within neutrino physics, which must be addressed with care. Here lies the importance of model building: at a fundamental level, the goal is to explore unanswered questions while complying with experimental observations. In this process, one must deal with several free parameters that arise from interactions and vacuum alignments. A specific vacuum alignment, for instance, may lead to a distinct neutrino mass matrix texture, but such a choice is rarely unique. This highlights the need to carefully examine whether a given alignment is truly special or simply one possibility among many. Furthermore, while discrete symmetries provide the scaffolding for the neutrino mass matrix, the final phenomenology depends equally on Yukawa couplings and vacuum alignments, an aspect often overlooked in the literature.  

The present work is therefore not merely concerned with assessing the predictability of a particular model, but with uncovering the range of possible textures that arise once freedom in parameter choices and alignments is systematically explored. By highlighting this perspective, we emphasize that what might appear as a single texture within a given framework can, upon closer inspection, conceal multiple viable realisations.  

It is in this spirit that we revisit our existing neutrino mass model based on the discrete group $\Delta(27)$. Our primary focus is not to propose an entirely new model, but rather to demonstrate how a simple generalization of the existing model, specifically, relaxing one of the vacuum alignments associated with the scalar fields opens up the possibility of constructing a rich variety of phenomenologically viable neutrino mass textures, while keeping the underlying symmetry structure intact. Some of these textures correspond to well-known cases, while others remain largely unexplored.  

Our analysis proceeds along both top-down and bottom-up directions. The top-down approach allows us to trace the origin of different textures from symmetry considerations, while the bottom-up perspective helps us embed phenomenologically consistent structures back into the model. In doing so, we demonstrate that even within a fixed theoretical framework, minimal modifications can unlock a diverse class of textures consistent with current experimental data, thereby providing new insights into neutrino phenomenology.  

The plan of the paper is as follows. In Section~\ref{THF}, we present the theoretical framework and introduce the general model. In Section~\ref{TextureRealisations}, we show how several well known neutrino mass textures emerge from the general model. Finally, in Section~\ref{SummaryConclusions}, we summarise our findings and present the conclusions.

\section{Theoretical Framework}
\label{THF}
The framework of the present model follows our earlier work Ref.~\cite{Dey:2023bfa}, where the SM gauge symmetry is extended by the discrete flavour group $\Delta(27)$, supplemented by $Z_3$\,\cite{Ma:2004yx,Hu:2006wk, Chakraborty:2024rgt} and $Z_{10}$\,\cite{Dey:2023rht, Dey:2024ctx, Chakraborty:2024eki, Dey:2025zld, Chakraborty:2024hhq, Chakraborty:2025juy, Goswami:2023eyy} symmetries. The model includes three heavy right handed neutrinos and a rich scalar sector containing $\Delta(27)$ triplet and singlet fields, whose charge assignments are chosen to control the Yukawa structures and forbid unwanted operators. After spontaneous symmetry breaking, with appropriate vacuum alignments of the scalar fields, both charged lepton and neutrino masses are generated. 

In the present work, we introduce a single modification to the model of Ref.~\cite{Dey:2023bfa}. While the original model assumes the vacuum alignment $\langle \Delta \rangle = v_\Delta(1,1,1)$ for the scalar triplet $\Delta$, here, we consider a more general form  $\langle \Delta \rangle = (v_{\Delta_1}, v_{\Delta_2}, v_{\Delta_3})$. Apart from this minor modification, all other aspects of the framework of Ref.~\cite{Dey:2023bfa} are kept unchanged.

With this generalised alignment, the Type-II seesaw mass matrix takes the form

\begin{eqnarray}
M_{II}&=& \frac{y_{\Delta}}{2}\begin{bmatrix}
  2 v_{\Delta_1} &v_{\Delta_3} &v_{\Delta_2}\\
  v_{\Delta_3}&2v_{\Delta_2}&v_{\Delta_1} \\
 v_{\Delta_2}&v_{\Delta_1}&2v_{\Delta_3} \\
 \end{bmatrix}
\end{eqnarray}
In a Type- I + II seesaw framework, the complex  Majorana neutrino mass matrix\,($M_{\nu}$) of our model adopts the following form
\begin{equation}
 M_{\nu} = 
 \begin{bmatrix}
 A + v_{\Delta_1}  y_{\Delta} & v_{\Delta_3}\,y_{\Delta}/2 & v_{\Delta_2}\, y_{\Delta}/2 \\
v_{\Delta_3}\,y_{\Delta}/2 & B +v_{\Delta_2} y_{\Delta} & C+ v_{\Delta_1}\,y_{\Delta}/2\\
v_{\Delta_2}\,y_{\Delta}/2& C+  v_{\Delta_1}\,y_{\Delta}/2  & D +v_{\Delta_3} y_{\Delta}  \\
 \end{bmatrix},
 \label{Neutrino Mass matrix}
\end{equation}
where,
\begin{align}
A = -\frac{v^2_h v^2_\chi y^2_1}{M_1 \Lambda^2},\quad
B =\frac{y_{r_2} v_{\kappa} v^2_h v^2_{\xi} v^2_{\chi} y^2_2}{(y^2_s v^2_\rho - y_{r_1}v_{\eta}y_{r_2}v_{\kappa}) \Lambda^4},\quad
C =-\frac{y_s v_\rho v^2_h v_\zeta v_\xi v^2_\chi y_2 y_3}{(y^2_s v^2_\rho - y_{r_1}v_{\eta}y_{r_2}v_{\kappa}) \Lambda^4},\quad
D =\frac{y_{r_1} v_{\eta} v^2_h v^2_{\zeta} v^2_{\chi} y^2_3}{(y^2_s v^2_\rho - y_{r_1}v_{\eta}y_{r_2}v_{\kappa}) \Lambda^4}.
\label{equations}
\end{align}

\section{Texture Realisations}
\label{TextureRealisations}
We generate a large set of numerical data points by matching the texture in Eq.~(\ref{Neutrino Mass matrix}) with a general complex symmetric neutrino mass matrix, which contains twelve parameters, three neutrino mass eigenvalues\,($m_1,\, m_2,\, m_3$), three mixing angles\,($\theta_{12},\, \theta_{13},\, \theta_{23}$), Dirac CP phase\,($\delta$), two Majorana phases\,($\alpha,\, \beta$), and three unphysical phases\,($\phi_1,\, \phi_2,\, \phi_3$). This allows us to systematically explore the texture structures that may be embedded in our model. 
The analysis is performed for both normal hierarchy (NH) and inverted hierarchy (IH). From the numerical scan, we select only those data points consistent with the experimental observables $\theta_{12},\, \theta_{13},\, \theta_{23},\, \Delta m^2_{21},\, \Delta m^2_{31},\, \delta$. We then perform scatter plots between different mass matrix elements to identify possible correlations, hierarchies, vanishing entries, or symmetries. 

\begin{figure}[h]
  \centering
    \subfigure[]{\includegraphics[width=0.24
  \textwidth]{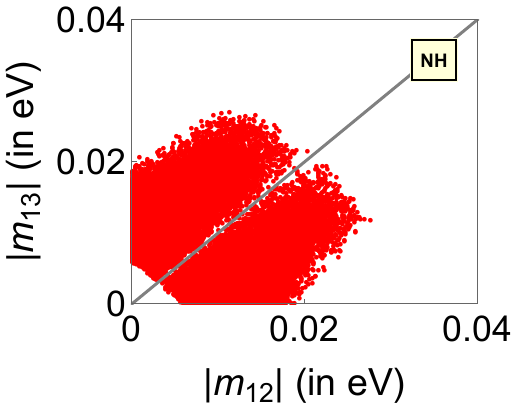}\label{fig:1(a)}}  
    \subfigure[]{\includegraphics[width=0.245\textwidth]{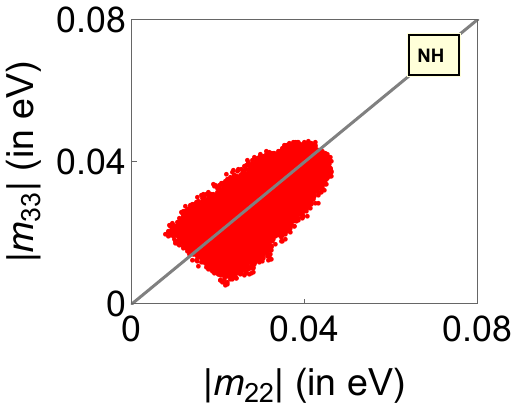}\label{fig:1(b)}}
    \subfigure[]{\includegraphics[width=0.245
    \textwidth]{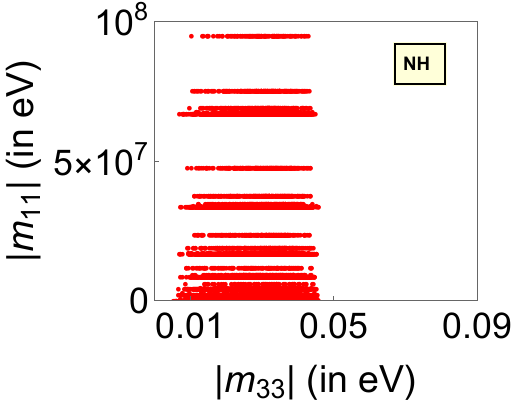}\label{fig:1(c)}}
    \subfigure[]{\includegraphics[width=0.245
    \textwidth]{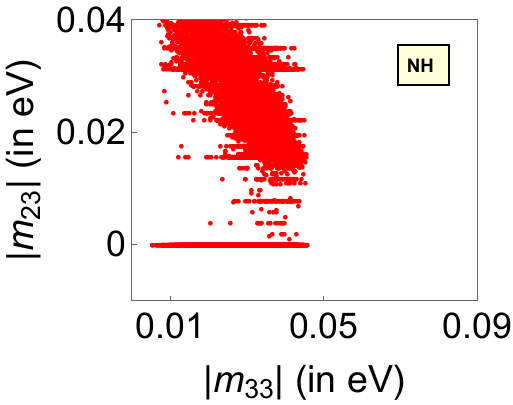}\label{fig:1(d)}}
\caption{Scatter plots illustrating possible correlations among neutrino mass matrix elements for NH}
\label{fig:1}
\end{figure}

\begin{figure}[h]
  \centering
    \subfigure[]{\includegraphics[width=0.24
  \textwidth]{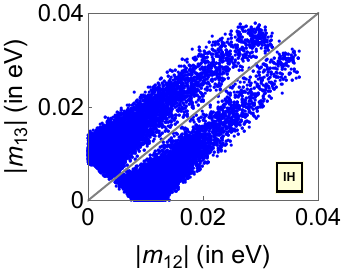}\label{fig:2(a)}} 
    \subfigure[]{\includegraphics[width=0.245\textwidth]{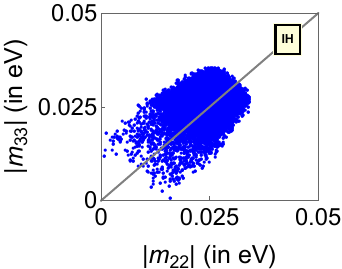}\label{fig:2(b)}}
    \subfigure[]{\includegraphics[width=0.245
    \textwidth]{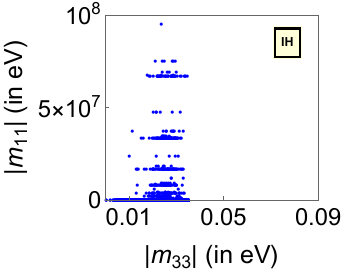}\label{fig:2(c)}}
    \subfigure[]{\includegraphics[width=0.245
    \textwidth]{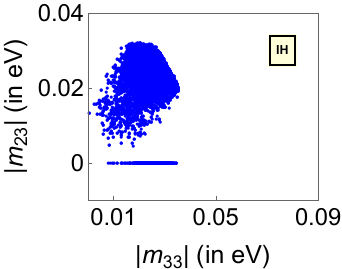}\label{fig:2(d)}}
\caption{Scatter plots illustrating possible correlations among neutrino mass matrix elements for IH}
\label{fig:2}
\end{figure}

From the scatter plots in Figure~\ref{fig:1} (for NH) and Figure~\ref{fig:2} (for IH), we observe the following hidden relations among the elements of $M_{\nu}$
\begin{eqnarray}
|m_{12}| \simeq   |m_{13}|, \quad
|m_{22}| =  |m_{33}|, \quad
|m_{11}| =  0, \quad
|m_{12}| =  0, \quad
|m_{13}| =  0, \quad
|m_{23}| =  0.
\label{elementzeroes}
\end{eqnarray}
We note from the scatter plots that while most of the matrix elements are typically of $\mathcal{O}(10^{-2})$, the element $|m_{11}|$ can reach values as large as $\mathcal{O}(10^{8})$ in certain regions of the parameter space.

	The relations in Eq.~(\ref{elementzeroes}) can appear in different combinations, and each consistent combination corresponds to a possible texture. Our aim is to systematically extract all such textures from the generated datasets and then examine whether the extracted textures can indeed be reproduced from the model itself, rather than being mere numerical coincidences. With this motivation, and assuming that the VEVs are real in our model while the Yukawa couplings are generally complex, we categorically discuss the possible textures in the following sections.
\subsection{Single Zero Textures}

From the datasets, we find a class of textures that contain one vanishing element in the neutrino mass matrix. The obtained single zero textures\,\cite{Lashin:2011dn} are 
\begin{equation}
|m_{11}|=0, \quad |m_{12}|=0, \quad |m_{13}|=0, \quad \text{and} \quad |m_{23}|=0
\nonumber
\end{equation}
Each of these structures can be realized within the present model through specific parameter choices or suitable vacuum alignments of the scalar triplet $\Delta$, as discussed below.
\begin{itemize}
\item[•]$|m_{11}| = 0$:
This texture emerges when the following condition, derived from Eq.~(\ref{Neutrino Mass matrix}), is satisfied
\begin{equation}
y_1 = \sqrt{y_{\Delta} v_{\Delta_1} M_1} \frac{\Lambda}{v_h v_\chi} 
\label{m11zero}
\end{equation}
\item[•]$|m_{12}| = 0$:
This texture can be achieved, when the scalar triplet $\Delta$ acquires the following alignment
\begin{equation}
\langle \Delta \rangle = (v_{\Delta_1}, v_{\Delta_2}, 0)
\label{m12zero}
\end{equation}
\item[•]$|m_{13}| = 0$:
This case follows when the vacuum alignment of $\Delta$ takes the form 
\begin{equation}
\langle \Delta \rangle = (v_{\Delta_1}, 0, v_{\Delta_3})
\label{m13zero}
\end{equation}
\item[•]$|m_{23}| = 0$:
At this step, it is important to note that achieving a particular texture of neutrino mass matrix does not solely depend on the choice of the discrete flavour symmetry groups and the proper choice of the vacuum alignments. In general, the neutrino mass matrix carries ample number of free parameters in terms of the Yukawa couplings, VEVs etc. Hence, we believe that tuning these parameters one can achieve the targeted neutrino mass matrix.
With this motivation, the texture with $|m_{23}| = 0$ can be obtained when the following relation among the parameters of Eq.~(\ref{Neutrino Mass matrix}) holds good
\begin{equation}
y_2 = \frac{v_{\Delta_1} y_{\Delta}}{2 y_3 y_s v_\rho v_\zeta v_\xi X},
\label{m23zero}
\end{equation}
where, $X = v^2_h v^2_{\chi}/(y^2_s v^2_\rho - y_{r_1}v_{\eta}y_{r_2}v_{\kappa}) \Lambda^4$.
\end{itemize}

\subsection{Two Zero Textures}
Having established the possible single zero textures, we identify a second category of textures characterized by two vanishing elements in the neutrino mass matrix. The viable two zero textures\,\cite{Fritzsch:2011qv} found from our analysis are
\begin{equation}
|m_{11}|=0,\,|m_{12}|=0;\quad |m_{11}|=0,\,|m_{13}|=0 \quad \text{and} \quad |m_{11}|=0,\,|m_{23}|=0 \nonumber
\end{equation}
All these structures can be realized from the mother texture in Eq.~(\ref{Neutrino Mass matrix}) when the corresponding conditions for single zero cases hold simultaneously

\begin{itemize}
\item[•]$|m_{11}| = 0$, $|m_{12}| = 0$:
This structure arises when both Eq.~(\ref{m11zero}) and Eq.~(\ref{m12zero}) are simultaneously true.

\item[•]$|m_{11}| = 0$, $|m_{13}| = 0$:
This structure arises when both Eq.~(\ref{m11zero}) and Eq.~(\ref{m13zero}) are simultaneously satisfied.

\item[•]$|m_{11}| = 0$, $|m_{23}| = 0$:
This structure arises when both Eq.~(\ref{m11zero}) and Eq.~(\ref{m23zero}) are simultaneously valid.
\end{itemize}

\subsection{Approximate Three Zero Textures}
We also identify two textures containing three vanishing elements in the neutrino mass matrix. In this case, matrix elements of $\mathcal{O}(10^{-5})$ are treated as effectively zero. With this approximation, the obtained three zero textures\,\cite{Zhang:2013mb} are 
\begin{equation}
|m_{11}|\sim 0,\,|m_{12}|\sim 0,\,|m_{23}|\sim 0 \quad \text{and} \quad |m_{11}|\sim 0,\,|m_{13}|\sim0,\,|m_{23}|\sim 0 \nonumber
\end{equation}

These configurations can naturally emerge within the framework when the corresponding single zero conditions are satisfied simultaneously.
\begin{itemize}
\item[•]$|m_{11}| \sim 0$, $|m_{12}| \sim 0$, $|m_{23}| \sim 0$:
This structure arises when Eq.~(\ref{m11zero}), Eq.~(\ref{m12zero}) and Eq.~(\ref{m23zero}) simultaneously holds good.

\item[•]$|m_{11}| \sim 0$, $|m_{13}| \sim 0$, $|m_{23}| \sim 0$:
This structure arises when Eq.~(\ref{m11zero}), Eq.~(\ref{m13zero}) and Eq.~(\ref{m23zero}) are simultaneously valid.
\end{itemize}

\subsection{The $\mu-\tau$ Mixed Symmetry}
\begin{figure}[h]
  \centering
    \subfigure[]{\includegraphics[width=0.243
  \textwidth]{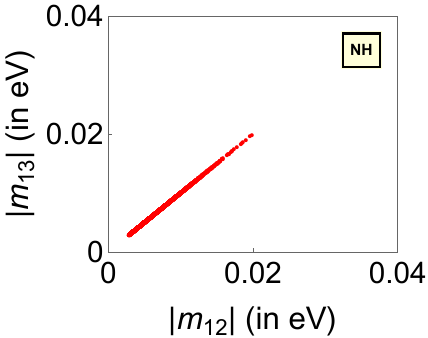}\label{fig:3(a)}}  
    \subfigure[]{\includegraphics[width=0.243\textwidth]{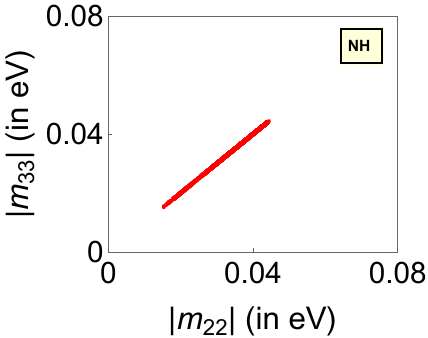}\label{fig:3(b)}}
    \subfigure[]{\includegraphics[width=0.243
    \textwidth]{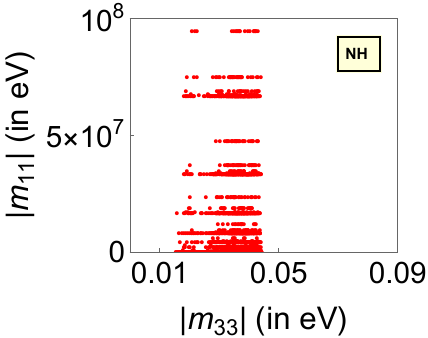}\label{fig:3(c)}}
    \subfigure[]{\includegraphics[width=0.243
    \textwidth]{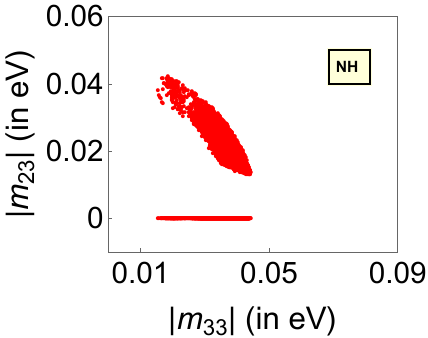}\label{fig:3(d)}}
\caption{Scatter plots illustrating the allowed parameter space that satisfies the $\mu$–$\tau$ mixed symmetry with two texture zeros in the case of NH}
\label{fig:3}
\end{figure}

\begin{figure}[h]
  \centering
    \subfigure[]{\includegraphics[width=0.243
  \textwidth]{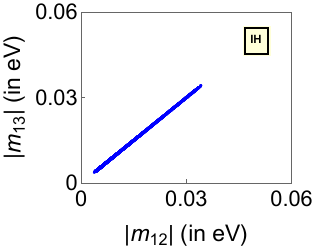}\label{fig:4(a)}} 
    \subfigure[]{\includegraphics[width=0.243\textwidth]{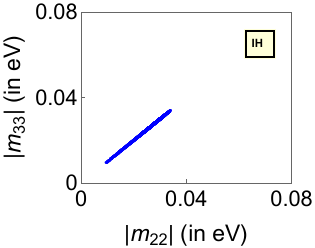}\label{fig:4(b)}}
    \subfigure[]{\includegraphics[width=0.243
    \textwidth]{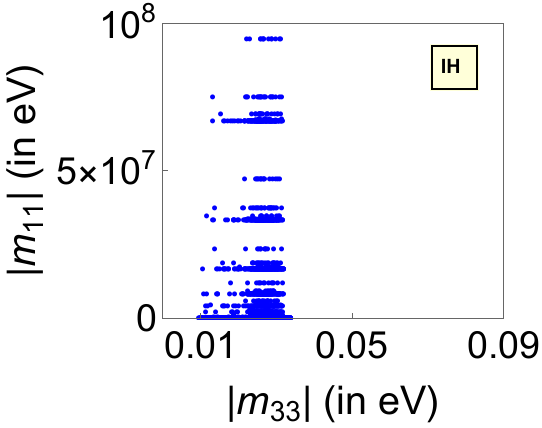}\label{fig:4(c)}}
    \subfigure[]{\includegraphics[width=0.243
    \textwidth]{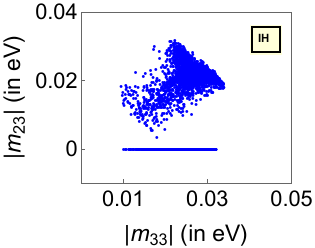}\label{fig:4(d)}}
\caption{Scatter plots illustrating the allowed parameter space that satisfies the $\mu$–$\tau$ mixed symmetry with two texture zeros in the case of IH}
\label{fig:4}
\end{figure}

The numerical analysis reveals that the mass matrix $m_{\nu}$ permits the relations $|m_{12}| \simeq |m_{13}|$ and $|m_{22}| = |m_{33}|$, it is therefore, interesting to explore whether the present framework can incorporate the well known $\mu-\tau$ mixed symmetry\,\cite{Dey:2022qpu}. The $\mu-\tau$ mixed symmetry posses the following relationship between the mass matrix elements
\begin{equation}
m_{13}= -m_{12}\quad \text{and}\quad  m_{33}= m_{22}^*
\end{equation}
We therefore filter those data sets which satisfy the $\mu-\tau$ mixed symmetry, and then we try to explore whether the said symmetry can be reconstructed from the theoretical framework or not. On filtering the data set we observe that, in addition to $\mu-\tau$ mixed symmetry, the present framework can shelter $\mu-\tau$ mixed symmetry with texture zeroes, i.e, $|m_{11}|= 0$, $|m_{23}|$= 0. To understand the said findings graphically, we highlight the scatter plots between the mass matrix elements for NH and IH in Figure\,\ref{fig:3} and Figure\,\ref{fig:4} respectively.
We now examine whether $\mu-\tau$ mixed symmetry can be reconstructed from our theoretical framework or not. As discussed previously, realisation of a particular neutrino texture is not determined solely by the imposed discrete flavour symmetry or the vacuum alignment pattern. Rather, the mass matrix depends on a rich set of free parameters, and appropriate interrelations among them can naturally reproduce specific structures, such as one zero, two zero, and approximate three zero textures. In this sense, the texture realisation serves as a phenomenological constraint on the underlying Lagrangian parameters. Guided by these considerations, we adopt the following parameter relations
\begin{equation}
v_{\Delta_2} = -v_{\Delta_3}\quad \text{and} \quad
y_{r_2}= \frac{2 v_{\Delta_3} y_{\Delta} + v_{\eta} v^2_{\zeta} y^*_{r_1} (y^2_3)^* X^*}{v_{\kappa} v^2_{\xi} y^2_2 X},
\label{mutauassumptions}
\end{equation}
For these choices, $M_{\nu}$ in Eq.~(\ref{Neutrino Mass matrix}) is simplified to
\begin{equation}
 M_{\nu} = 
 \begin{bmatrix}
 A + v_{\Delta_1}  y_{\Delta} & v_{\Delta_3}\,y_{\Delta}/2 & -v_{\Delta_3}\, y_{\Delta}/2 \\
v_{\Delta_3}\,y_{\Delta}/2 & D^* +v_{\Delta_3} y_{\Delta}^* & C+ v_{\Delta_1}\,y_{\Delta}/2\\
v_{\Delta_2}\,y_{\Delta}/2& C+  v_{\Delta_1}\,y_{\Delta}/2  & D +v_{\Delta_3} y_{\Delta}  \\
 \end{bmatrix}.
 \label{Neutrino Mass matrix2}
\end{equation}
This structure of $M_{\nu}$ resonates with the $\mu-\tau$ mixed symmetry. 	
	It is also possible to reconstruct $\mu-\tau$ mixed symmetry with texture zeroes. For this case, in addition to the choices mentioned in Eq.\,(\ref{mutauassumptions}), the Eq.~(\ref{m11zero}) and Eq.~(\ref{m23zero}) are simultaneously valid.
	
	In this context, we note that under the assumption of real VEVs and complex Yukawa couplings, the other variants of the $\mu$-$\tau$ mixed symmetry\,\cite{Chakraborty:2023msb} cannot be realised from the mother texture in Eq.~(\ref{Neutrino Mass matrix}).

\section{Summary and Conclusions}
\label{SummaryConclusions}
In this work, we have considered a previously proposed framework based on the $\Delta(27)$ discrete symmetry and additional cyclic symmetries. We have demonstrated that by generalising only the vacuum alignment of the scalar triplet $\Delta$ from the previous model, it is possible to realise a wide variety of neutrino mass textures within a single framework. By systematically scanning the parameter space, we have shown that the model can naturally accommodate single zero, two zero, and approximate three zero textures, as well as the $\mu$-$\tau$ mixed symmetry. These textures emerge from specific relations among the model parameters and vacuum expectation values within the general mother texture. A detailed phenomenological study of the predictions for neutrino masses, mixing angles, and CP phases lies outside the scope of this work and is subject to future investigations. Overall, this study provides a unified theoretical origin for a broad class of neutrino mass textures discussed in the literature.

\bibliographystyle{unsrt}
\setlength{\bibsep}{0.1em} 
\bibliography{reference1.bib}
\end{document}